# Molecular echoes in space and time

Kang Lin[1], Peifen Lu[1], Junyang Ma[1], Xiaochun Gong[1], Qiying
Song[1], Qinying Ji[1], Wenbin Zhang[1], Heping Zeng[1]*, and Jian Wu[1]†
[1]*State Key Laboratory of Precision Spectroscopy,
East China Normal University, Shanghai 200062, China*

Gabriel Karras[2], Guillaume Siour[3], Jean-Michel Hartmann[3], and Olivier Faucher[2]‡
[2]*Laboratoire Interdisciplinaire CARNOT de Bourgogne,
UMR 6303 CNRS-Université Bourgogne Franche-Comté, BP 47870, 21078 Dijon, France and*
[3]*Laboratoire Interuniversitaire des Systèmes Atmosphériques (LISA) CNRS (UMR 7583),
Université Paris Est Créteil, Université Paris Diderot, Institut Pierre-Simon Laplace,
Université Paris Est Créteil, 94010 Créteil Cedex, France*

Erez Gershnabel[4], Yehiam Prior[4]§, and Ilya Sh. Averbukh[4]¶
[4]*AMOS and Department of Chemical Physics,
Weizmann Institute of Science, Rehovot 76100, Israel*

(Dated: June 27, 2016)

**Mountain echoes are a well-known phenomenon, where an impulse excitation is mirrored by the rocks to generate a replica of the original stimulus, often with reverberating recurrences. For spin echoes in magnetic resonance and photon echoes in atomic and molecular systems the role of the mirror is played by a second, time delayed pulse which is able to reverse the flow of time and recreate the original event. Recently, laser-induced rotational alignment and orientation echoes were introduced for molecular gases, and discussed in terms of rotational-phase-space filamentation. Here we present, for the first time, a direct spatiotemporal analysis of various molecular alignment echoes by means of coincidence Coulomb explosion imaging. We observe hitherto unreported spatially rotated echoes, that depend on the polarization direction of the pump pulses, and find surprising imaginary echoes at negative times.**

PACS numbers:

*hpzeng@phy.ecnu.edu.cn
†jwu@phy.ecnu.edu.cn
‡olivier.faucher@u-bourgogne.fr
§yehiam.prior@weizmann.ac.il
¶ilya.averbukh@weizmann.ac.il

In 1950, Erwin Hahn reported [1] that if a spin system is irradiated by two properly timed and shaped pulses, a third pulse appears at twice the delay between the first two. The intuitive explanation was given in terms of time reversal, namely the second pulse reverses the direction of propagation of the original excitation, leading to reappearance of the original impulse [2]. In the absence of interaction with the environment, the full original excitation is recovered, but with environmental influences, various dephasing and energy loss processes may be probed. Following the original discovery in the realms of spins, echoes were observed in many other nonlinear physical situations such as photon echo [3], cyclotron echo [4], plasma wave echo [5], echoes in cold atoms [6, 7], cavity QED [8], and even in particle accelerators [9, 10]. Echoes form the basis for many modern methodologies ranging from Magnetic Resonance Imaging (MRI) [11] to short-wavelength radiation generation in free-electron lasers [12–15]. Echoes are a classical phenomenon that is different from another well-known effect: quantum revivals [16–18] which are caused by the energy quantization of physical systems. Recently, a new type of echoes was introduced: molecular alignment echoes [19, 20]. When a gas of molecules undergoes excitation by an ultrashort laser pulse, the molecules experience a torque causing transient alignment of the ensemble along the laser polarization axis (for a review of laser molecular alignment, see [21–24]). A pair of time-delayed laser pulses results in three alignment events: two of them immediately following each excitation, and a third one, an echo, that appears with a delay equal to that between the exciting pulses. This delay can be shorter than the rotational revival time, so that the echo provides access to rapidly dephasing molecular dynamics. The formation of these echoes is caused by the kick-induced filamentation of the rotational phase space [19], a phenomenon well known in the physics of particle accelerators [32]. Moreover, fractional echoes were predicted and observed in molecular alignment [20], at times which are rational fractions of the delay between the pulses. While the primary (full) alignment echo was observed by measuring the laser-induced birefringence [19], for the fractional echoes, higher order moments of the molecular angular distribution are needed. Optical detection of the lowest order (1/2) fractional echo demanded measurement of the third harmonic of a probe pulse interrogating the kicked molecules [20], and even higher harmonics are required for the more complex higher order fractional echoes. This requirement limits the feasibility of using purely optical techniques to probe the details of the rotational echo process.

In the present work, we overcome this obstacle by providing direct access to the spatiotemporal molecular dynamics with femtosecond time-resolution, by means of the Cold Target Recoil Ion Momentum Spectroscopy (COLTRIMS) technique [25]. This method provides an invaluable addition to the toolbox for diagnostics of molecular evolution. Recently, this approach was successfully applied to visualizing molecular unidirectional rotation [26], with similar results reported by the Ohshima group[27]. Our current experimental arrangement is shown in Fig. 1, with details given in the Methods section below and in [26]. Here, as described above, the molecular excitation is achieved using two laser pulses. At times when the system is to be probed, a strong circularly polarized pulse is used to Coulomb-explode the molecules. The detected flying fragments provide a signature of the spatial orientation of the molecule at the time it was exploded. Our COLTRIMS setup enables us to spatiotemporally observe full and fractional echoes of high fractional order for the first time. Building upon the angular resolution provided by COLTRIMS, we predict and observe "rotated echoes", which are alignment echoes whose angular orientation is controlled by the polarization directions of the first two pulses. We further predict "imaginary alignment echoes" which should appear at negative time delays. This counter-intuitive phenomenon is explained by analytically continuing the rotational dynamics induced by the two pulses to negative evolution times. Clearly, such a "time machine" regime is not possible in classical physics, but the phenomenon of quantum rotational revivals provides a unique opportunity to study echoes at negative time, and here we report the first experimental observation of the imaginary echoes.

In what follows, we start with a discussion of the spatio-temporal dynamics of the various echoes. We then report the COLTRIMS-enabled experimental observation of the full, fractional (1/2 and 1/3), rotated and imaginary echoes. All observations are compared with classical and quantum mechanical simulations, within the 2D and 3D frameworks.

### Dynamics of echo formation

Consider an ensemble of symmetric linear molecular rotors uniformly dispersed in angle $\theta$, and with a Gaussian distribution spread ($\sigma$) in angular velocity [19, 20]. At time $t = 0$, a short, nonresonant linearly polarized laser pulse (delta-kick) is applied to the gas. The interaction potential is $V(\theta, t) = -(\Delta\alpha/4)E^2(t)\cos^2\theta$ [28, 29], where $\Delta\alpha$ is the polarizability anisotropy, and $E(t)$ is the envelope of the laser pulse. As is well known, such an exciting pulse leads to molecular alignment along the laser field polarization direction ($\theta = 0$) [21–24]. The ensemble averaged degree of alignment $\langle \cos^2\theta \rangle$ is typically referred to as the alignment factor. Following the kick, the angle $\theta$ of a rotor and its angular velocity are given by:

$$\theta = \theta_0 + \omega t, \quad \omega = \omega_0 - \Omega \sin(2\theta_0) .\qquad(1)$$

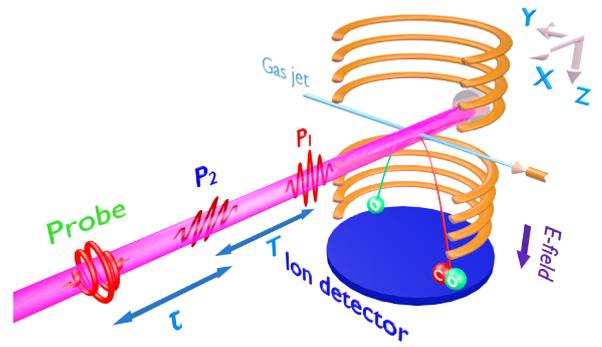

FIG. 1: **Experimental setup**. A supersonic gas jet of $CO_2$ molecules subject to a pair of polarization-skewed and time-delayed femtosecond laser pulses in an ultrahigh vacuum chamber of COLTRIMS is used to create molecular alignment echoes. By scanning the delay of an intense circularly polarized probe pulse, snapshots of the angular alignment of the molecular axis at various times are taken via Coulomb explosion.

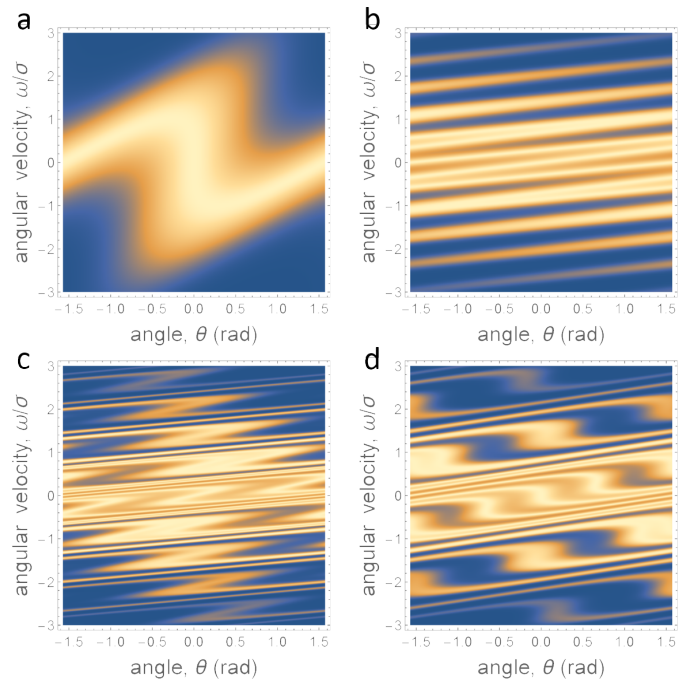

FIG. 2: **Filamentation of the phase space density distribution.** For all parts of this figure $\Omega/\sigma = 1$. **a)** Shortly after the kick, $\sigma t = 0.5$, the density distribution folds, resulting in a transient alignment along the direction $\theta = 0$. The folded pattern is centrally-symmetric with respect to the phase-space point $(\theta, \omega) = (0, 0)$ that is not affected by the kick. **b)** On longer time scale, $\sigma t = 5$, the probability density becomes wrinkled, and it develops multiple parallel filaments. **c)** After a second kick is applied at $t = T$ (with $\sigma T = 5$ and $\Omega_2/\Omega_1 = 1/3$), every filament in (b) folds in a manner similar to (a) giving rise to an alignment echo near $\theta = 0$ at time $\tau \sim T$ after the second kick ($\sigma\tau = 4.56$). **d)** At $\tau \approx T/2$ ($\sigma\tau = 2.293$), a fractional echo is formed.

Here $\omega_0, \theta_0$ are the initial conditions, and $\Omega$ is proportional to the intensity of the kick. The transformation (1) is equivalent to the Chirikov-Taylor map used in studies of deterministic chaos [30]. After the pulse, the phase space probability distribution is given by [19]:

$$f(\omega, \theta, t) = \frac{1}{2\pi}\frac{1}{\sqrt{2\pi}\sigma}\exp\left[-\frac{[\omega - \Omega\sin(2\omega t - 2\theta)]^2}{2\sigma^2}\right].\qquad(2)$$

Figures 2a,b show the evolution of the initial distribution with time, from (a) a single centrally-symmetric folded pattern appearing shortly after the kick, to (b) multiple parallel filaments emerging on longer time scales, when the alignment signal $\langle \cos^2\theta \rangle(t)$ diminishes. With time, the number of these filaments grows, and since the phase space volume is constant, their width narrows, till they become almost horizontal and uniform in density (Fig. 2b). The neighboring filaments are separated in angular velocity by $\sim \pi/t$, as can be seen from Eq.(2). Such a filamentation of phase space is well known in stellar systems [31] and in accelerator physics [32], and the folding of the phase space distribution shown in (Fig. 2a) has much in common with the bunching effect observed in particle accelerators [9, 10].

At $t = T$, the filamented ensemble is subject to a second pulse with the same linear polarization. With time, every filament in Fig. 2b folds in a manner similar to Fig. 2a. Due to the "quantization" of the angular velocities, at time $\tau \sim T$ after the second kick (Fig.2c), the folded parts bunch up near $\theta = 0$, resulting in an alignment echo. In a similar manner, higher order echoes are also formed at delays $2T, 3T, ....$. Moreover, as shown in [20], at rational fractions of the delay time (such as $\tau = T/2, T/3, ...$), some highly symmetric structures in the phase space appear due to synchronization of the folded features from non-neighboring filaments (Fig. 2d). An analytical expression for the time-dependent moments $\langle \cos(2n\theta) \rangle$, related to the expression derived in [20], completely supports these geometric arguments:

$$\langle \cos(2n\theta) \rangle(\tau) =$$
$$\sum_{k=-\infty}^{k=\infty}(-1)^k e^{-2\sigma^2(n\tau-kT)^2}J_{k+n}[2n\Omega_2\tau]J_k[2n\Omega_1(n\tau-kT)].\qquad(3)$$

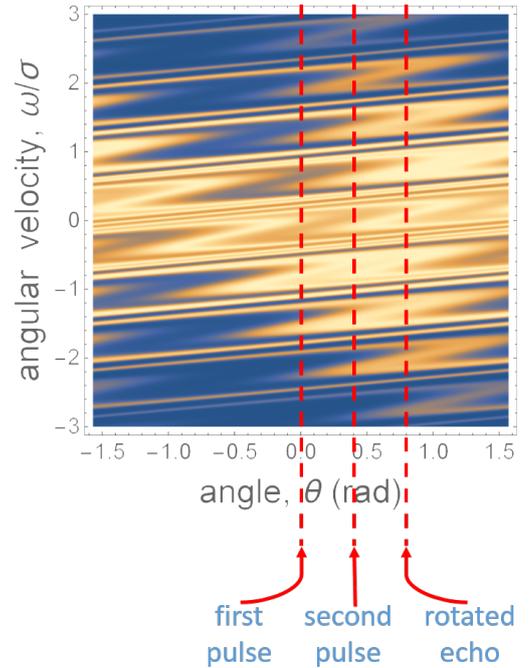

FIG. 3: **Formation of the rotated echo.** The first pulse is applied at $t = 0$, $\theta = 0$, $\Omega_1/\sigma = 1$, $\sigma T = 5$, a second pulse is applied at $t = T(4.56)$ ps, crossing angle $\beta = 0.4$ with respect to the first pulse, $\Omega_2/\Omega_1 = 1/3$; The rotated full echo is formed at $\sigma\tau = 4.56(\approx \sigma T)$, and angle $\theta = 0.8$ ($2\beta$).

Here $J_m(z)$ is the $m$-th order Bessel function of the first kind. For $\sigma T \gg 1$ only positive $k$ should be taken into account. The equation describes a sequence of echoes at $\tau = \frac{k}{n}T$ where $k$ is an integer. Here $n = 1$ gives rise to the primary (full) alignment echoes, whereas for $n > 1$, the equation describes the series of "fractional echoes".

The above phase-space analysis has a considerable power, enabling us to predict two other echo phenomena - rotated and imaginary echoes. As discussed in [20], for optical observation of the fractional echoes, higher moments $\langle \cos(2n\theta) \rangle$ ($n > 1$) should be measured, which is not a trivial task. An alternative approach adopted in the present work uses Coulomb explosion imaging of the molecular angular distribution for a direct visualization of the fractional echoes.

**Rotated echoes.** Let us now consider the case where the second kick is applied (at $t = T$) polarized at a crossing angle $\beta$ with respect to the first one. This pulse will fold each filament in a way similar to Fig. 2a, however the initial position of the symmetry center of the folded features will be at $(\theta, \omega) = (\beta, m\pi/T + \beta/T)$, where $m$ is the filament number. The folded parts of the filaments will bunch up after another delay $\tau = T$ at the angular position $\theta = [\beta + (m\pi/T + \beta/T)]T\,(\text{mod}\,\pi) = 2\beta$. Thus, the echo will be not only delayed in time, but also rotated by an additional angle $\beta$ with respect to the second pulse, or $2\beta$ with respect to the first one ( Fig. 3). Here we report the first experimental observation of these rotated

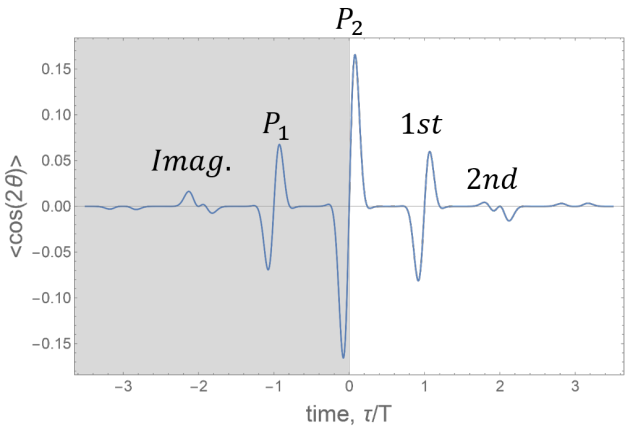

FIG. 4: **Echoes for positive and negative time** Two pulses are applied at $\tau/T =$-1 and at $\tau = 0$. Real echoes are seen at $\tau/T =$1,2,3 after the second pulse, and imaginary echoes are visible at $\tau/T =$-2,-3, i.e. at negative propagation time. The plotted quantity, $\langle\cos(2\theta)\rangle$ (Equ. 3 with n=1), is related to the standard alignment factor $\langle\cos^2\theta\rangle$ by the standard trigonometric identity $\cos(2\alpha) = 2\cos^2\alpha - 1$.

echoes (see the results section below).

**Imaginary echoes.** All the echoes discussed so far appear in the course of free evolution after the second pulse kicks the filamented phase-space pattern prepared by the first one. If, after the second kick, we were able to travel "back in time", the same qualitative arguments would predict the appearance of echoes at negative times - clearly an impossibility in real physical systems. Fortunately, the phenomenon of quantum revivals [16–18] provides such an opportunity to probe "negative" times. The system dynamics just before the revival time corresponds to the classical evolution before $\tau = 0$. Thus we predict a series of echoes just before the revival of the first pump pulse. The shape of these echoes may be obtained from Eq.(3) by extending it to negative $\tau$. We term them "imaginary echoes" in line with refs. [33, 34] where related phenomena were first considered. Fig. 4 presents echoes and imaginary echoes derived from Eq.(3) in a unified way.

### Experimental results

**Full and fractional echoes.** All the experimental results in this paper were obtained in our COLRTIMS set-up, shown in Fig. 1, with details given in the Methods section below and in Ref. [26]. The raw data out of a COLTRIMS experiment are depicted as a "carpet" of probability distribution where the horizontal axis is the probe time delay, $\tau$, measured from the second pulse, the vertical axis is the angle, $\phi$, in the Y-Z plane away from the Z direction and the density is given by the color code. To increase the visibility and eliminate the bias induced by possible imperfections in the circularity of the probe pulse, we normalize the total probability of the angular distribution to unity for each time delay and then sub-tract the averaged angular distribution between the two pump pulses. We start with the primary (full) alignment echo. In Fig. 5, the first pulse ($P_1$) arrives at $\tau = -3$ ps and leads to the impulsive alignment of the molecular axis along the polarization direction of $P_1$, which vanishes with time due to the dispersion of molecular rotational velocities. A second pump ($P_2$) polarized parallel to $P_1$ is applied at $\tau = 0$ ps, and leads to another immediate response. As discussed above, the following free rotational dynamics results in an alignment echo centered at twice the time delay after the first pulse, i.e. at $\tau = 3$ ps. This echo is seen between the dashed lines in Fig. 5a. The corresponding polar plot of the angular distribution around $\tau = 2.5$ ps is shown in Fig. 5b, demonstrating clear alignment along the same polarization direction as that of the exciting pulses. In previous optical studies of the alignment process after a double pulse excitation [19] (see also the related measurements of Ref. [35]), the ensemble averaged alignment signal, $\langle\cos^2\theta\rangle$ was measured, while here (with the COLTRIMS methodology) we directly visualize the angular distribution of the molecules. As is clearly seen from the polar plot around $\tau = 3.2$ ps (Fig. 5c), the echo not only shows alignment, but also a rapid transition from the alignment to anti-alignment. The red lines in Figs. 5b,c are the results of numerical simulations showing very good agreement with the measured signals. In the supplementary material we discuss both quantum-mechanical and classical simulations (see Figs. S1 a-c and Figs. S2 a-c, respectively), and the two agree very well with one another and with the experimental results.

Besides the primary (full) echoes at $\tau$=T, and the higher order full echoes at $\tau$=2T, 3T... (not shown), fractional echoes can now be clearly seen - Figs. 5d-i depict the situation for the one half and one third echoes at $\tau = T/2$ and $T/3$. To identify the temporal evolution of the fractional echoes, we increased the time delay between $P_1$ and $P_2$ to 5 ps. The measured time-dependent angular distribution of the 1/2 fractional echo (emerging around $\tau = T/2$) is shown in Figs. 5d-f. Whereas the full echo is cigar shaped, the fractional 1/2 echo exhibits much more complex structures of a cross shape at $\tau = 2.2$ ps followed by a butterfly distribution at $\tau = 2.7$ ps, as shown in Figs. 5e and f. In Figs. 5g-i, we also visualize the 1/3 echo. It emerges around $\tau = T/3 \sim 1.66$ ps for $T = 5$ ps, and shows a more elaborate six-lobe molecular angular distributions. For both the fractional 1/2 and 1/3 echoes, the experimental results shown in the polar plots are matched very well by the simulations (red in in these figures), and here, too, the quantum-mechanical and classical simulations reproduce each other very well, as is discussed in detail in the supplementary material (see Supplementary material Figs. S1 g-i and Figs. S2 g-i, respectively).

**Rotated and imaginary echoes** A rotated echo that is caused by a time-delayed second pump pulse that is polarized at a crossing angle $\beta = -20°$ with respect to the first one was observed. The measured angular distribution and its quantum and classical simulations are presented in Figs. 6a-f. The rotated full echo emerges around $\tau = 3$ ps with the maximum of its (experimentally broadened) angular distribution first at -40°(alignment) followed by a rapid transfer to -130°(anti-alignment). As predicted above, the echo is rotated by $2\beta$ with respect to the first pulse. This angular dependence of the rotated echo was further verified (not shown) by varying $\beta$ over a range of values, and observing the corresponding angular dependence of the echo. These measurements provide the first spatiotemporal visualization of rotated echoes of impulsively aligned molecules. The entire simulated time evolution of the rotated echoes is described in the supplementary movie.

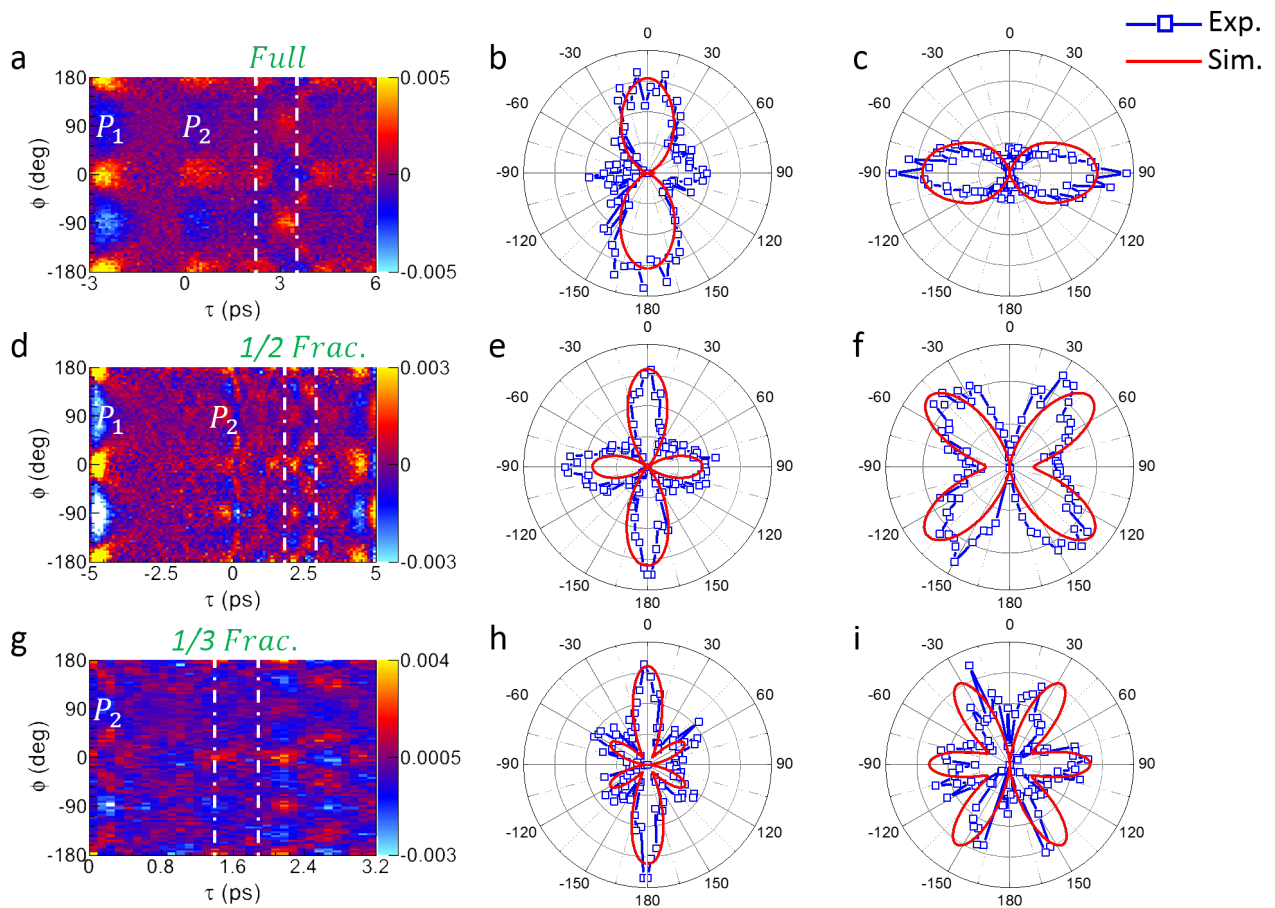

FIG. 5: **Time-dependent angular distributions of full, 1/2 and 1/3 echoes, experimentally measured and theoretically simulated for $CO_2$. a)** Carpet of the angular distribution of full echo for parallel input pulses, $P_1$ and $P_2$, with $T = 3$ ps time delay between them. **b)** and **c)** Polar plots of the alignment and anti-alignment regions of the distribution for the full echo, showing vertical and horizontal directionalities respectively. **d-f)** The same, for the 1/2 fractional echo, with $T = 5$ ps time delay. **g-i)** The same for 1/3 fractional echoes with $T = 5$ ps time delay. In all cases, the solid red line is the quantum mechanically simulated distribution.

The preceding results show that the time evolution of the twice-kicked ensemble gives rise to echoes that emerge **after** the application of the second laser pulse. Eq.(3) predicts imaginary echoes for "negative" times **before** the first pulse as depicted in Fig. 4. The time domain near (and earlier than) quantum full revival of the rota-tional excitation offers a unique opportunity to observe imaginary echoes at "negative" times.

Figure 7 shows: (a) the time-dependent angular distribution and (b) the alignment factor for a double-pulse excited gas of $CO_2$ molecules measured for long waiting times after the excitation. The red and blue peaks are the full quantum revival of the two pump pulses, the green signal is the quantum revival of the first full echo and the purple peak around 37 ps is the imaginary echo that precedes the revival of the first pulse by exactly the delay between the two exciting pulses.

$N_2O$ is a different molecule, with the rotational constant such that its revival time of 39.9 ps is very similar to that of the $CO_2$ molecule. Unlike $CO_2$, due to different symmetry, this molecule does not demonstrate quarter revivals, and therefore the time window before the full rotational revival provides more room for the observation of the imaginary echo. Fig. 8 depicts a measurement near 32-44 ps in $N_2O$. The replicas of the first and second pulses are clearly seen in the figure, both in the carpet (Fig. 8a) and in the angular alignment $\langle \cos^2 \phi \rangle(t)$ (Fig. 8b). At earlier times, the imaginary echo is seen (marked by the two vertical lines), completely isolated from any neighboring signals. The polar plot of this imaginary echo is depicted in Fig. 8c.

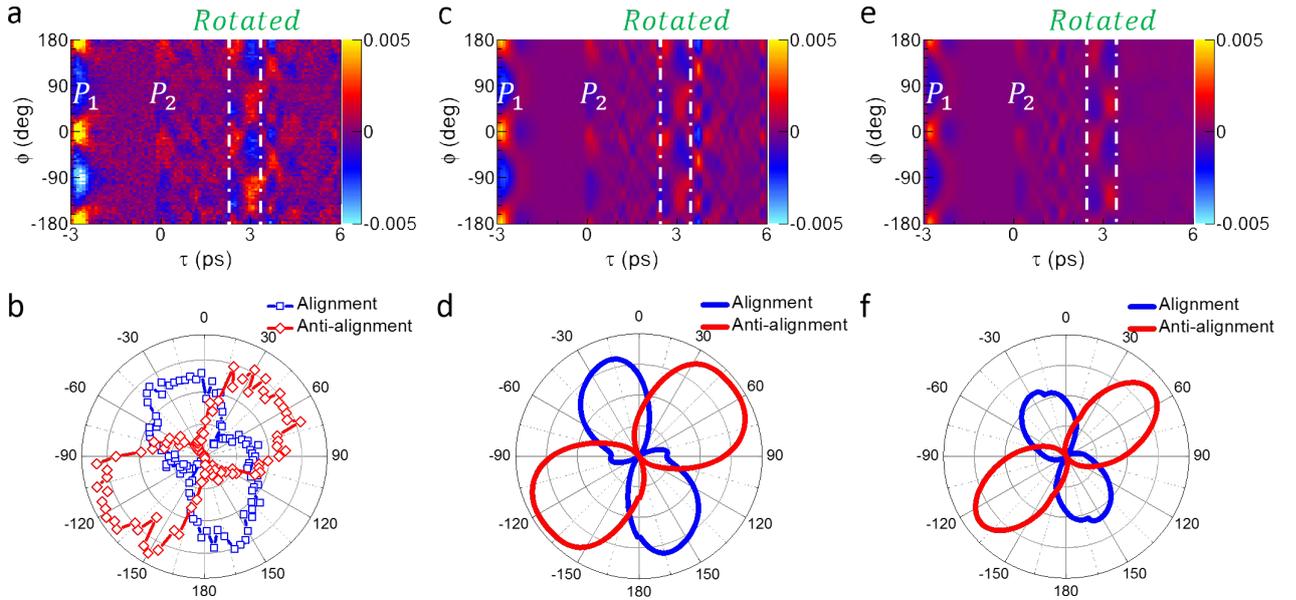

FIG. 6: **Experimentally measured and theoretically simulated time-dependent angular distributions of the rotated echo in $CO_2$.** $P_1$ and $P_2$ arrive at -3 ps and 0 ps respectively, with a crossing angle of $\beta = -20°$. **a)** Angular distribution of the rotated echo as a function of the prob time delay. **b)** Polar plots of the rotated echo at 3 ps. **c,d)** quantum and **e,f)** classical simulations.

### Summary

By using coincidence Coulomb explosion imaging, we study the spatiotemporal dynamics of various alignment echoes. In addition to the full and fractional echoes, we report new types of echoes: rotated echoes and imaginary echoes. When a molecule is impulsively kicked by a pair of time-delayed and nonidentically polarized ultrashort laser pulses, the alignment echo appears at twice the time delay between the pumps, polarized along a direction that is twice the angle between them. We predict these echoes, calculate them classically and quantum mechanically, and observe them experimentally. Furthermore, if in the equation of motion for the echo formation time is allowed to propagate backwards, imaginary echoes are predicted at negative time before the first pulse. To overcome the physical impossibility of working in negative time, we probe a time window just prior to the full quantum revival, which is fully equivalent to the window just prior to the origin. Imaginary echoes are indeed observed and analyzed.

### Methods

The alignment echoes are generated in a molecular beam subject to a pair of femtosecond laser pulses, accompanied by an intense circularly polarized probe pulse that Coulomb-explodes the molecules to image their orientation at various time slices, as schematically shown in Fig. 1. The three pulses are derived from a linearly polarized femtosecond laser pulse (25 fs, 790 nm, 10 kHz, Ti:sapphire Femtolasers) and are denoted as $P_1$, $P_2$, and probe pulse respectively. Two individual half wave plates are placed in the optical path of the pulses to polarize $P_1$ along the Z axis, and to adjust the polarization of $P_2$ for the different experiments. The three pulses are focused onto a supersonic molecular beam in an ultrahigh vacuum chamber of the COLTRIMS apparatus by a concave silver mirror (f = 7.5 cm). The temporal duration of each pulse in the interaction region is estimated to be ~60 fs by tracing the time-delay-dependent single-ionization yield as a cross-correlation of every two pulses. To achieve strong echo signals, the intensity ratio of $I_2/I_1$ in the case of parallel polarizations of $P_2$ and $P_1$ is experimentally optimized to around 0.3. The intensities of $P_1$, $P_2$ in the reaction area are measured to be $I_1 \sim 5.0 \times 10^{13}$ W/cm$^2$, $I_2 \sim 1.4 \times 10^{13}$ W/cm$^2$ and the intensity of the probe pulse is $\sim 4 \times 10^{14}$ W/cm$^2$. The produced fragment ions are accelerated and guided by a weak homogeneous static electric field (~20 V/cm) and then detected by a time- and position-sensitive microchannel plate detector. In general, the three-dimensional momenta of the ions are retrieved from the measured times-of-flights and positions-of-impacts. In our case of linear molecules, the direction of the molecular axis is retrieved from the double-ionization-induced Coulomb explosion channel of $[CO_2 + n\hbar\omega \rightarrow CO^+ + O^+ + 2e]$ and $[N_2O + n\hbar\omega \rightarrow NO^+ +$

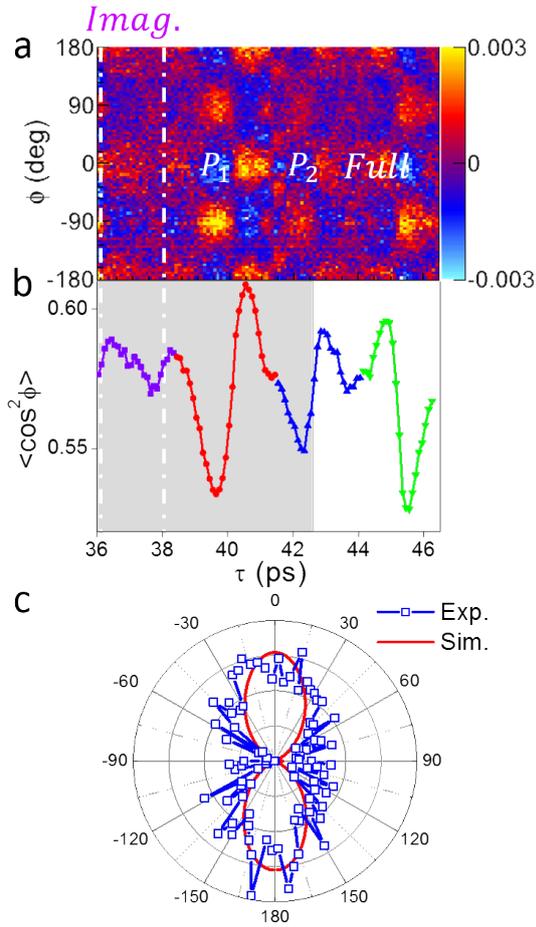

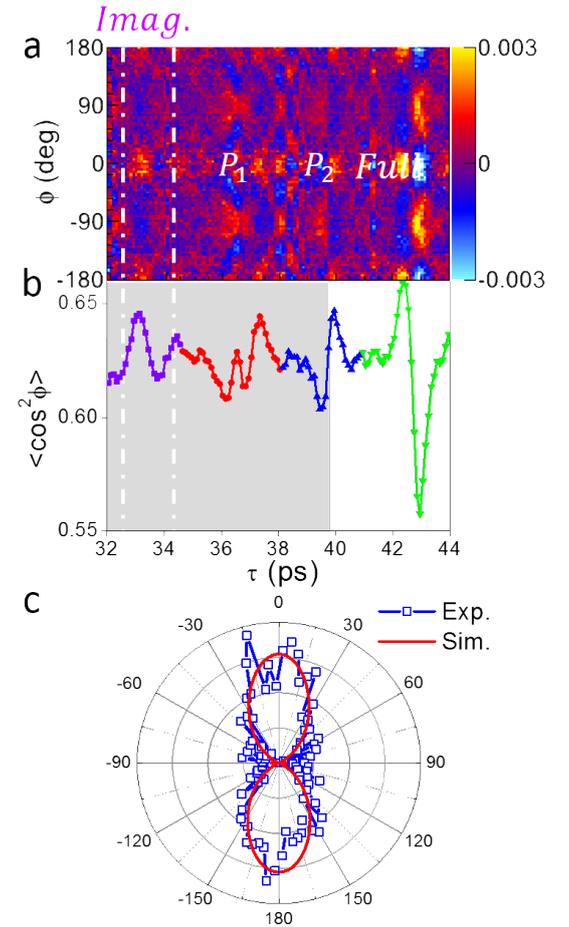

FIG. 7: **Experimentally measured and theoretically simulated time-dependent angular distributions of the imaginary echo in CO$_2$.** $P_1$ and $P_2$ arrive at -2.6 ps and 0 ps respectively. **a)** A "carpet" describing the angular distribution of the imaginary echo as a function of the probe time delay. **b)** The angular alignment, $\langle \cos^2 \phi \rangle$, as a function of the probe delay (red - first pulse, blue - second pulse, green - first full echo, purple - imaginary echo). **c)** Polar plots of the imaginary echo around 36.6 ps (blue - experiment, red - simulation).

FIG. 8: **Experimentally measured and theoretically simulated time-dependent angular distributions of the imaginary echo in N$_2$O.** $P_1$ and $P_2$ arrive at -3 ps and 0 ps respectively. **a)** A "carpet" describing the angular distribution of the imaginary echo as a function of the probe time delay. **b)** The angular alignment, $\langle \cos^2 \phi \rangle$, as a function of the probe delay (red - first pulse, blue - second pulse, green - first full echo, purple - imaginary echo). **c)** Polar plots of the imaginary echo around 33 ps (blue - experiment, red - simulation).

N$^+$ + 2e] for CO$_2$ and N$_2$O respectively. The angle $\phi$ away from the Z axis (in the Y-Z plane) is tracked as a function of probe pulse delay to reveal the evolution of the molecular orientation. The probe pulse is circularly polarized in the y-z plane, and we restrict our data analysis in this plane by selecting molecules confined to [-20°, 20°] with respect to the Y-Z plane (see Fig. 1).

[1] Hahn, E. L. Spin Echoes. *Phys. Rev.* **80**, 589 (1950).

[2] Hahn, E. L. Free nuclear induction. *Physics Today* **6**(11), 4 (1953).

[3] Kurnit, N. A., Abella, I. D. & Hartmann, S. R. Observation of a Photon Echo. *Phys. Rev. Lett.* **13**, 567 (1964).

[4] Hill, R. M. & Kaplan, D .E. Cyclotron Resonance Echo. *Phys. Rev. Lett.* **14**, 1062 (1965).

[5] Gould, R. W., O'Neil, T. M. & Malmberg, J. H. Plasma Wave Echo. *Phys. Rev. Lett.* **19**, 219 (1967).

[6] Bulatov, A., Kuklov, A., Vugmeister, B. E. & Rabitz, H. Echo in optical lattices: Stimulated revival of breathing oscillations. *Phys. Rev. A* **57**, 3788 (1998).

[7] Herrera, M., Antonsen, T. M., Ott, E. & Fishman, S. Echoes and revival echoes in systems of anharmonically confined atoms. *Phys. Rev. A* **86**, 023613 (2012).

[8] Meunier, T., Gleyzes, S., Maioli, P., Auffeves, A., Nogues, G., Brune, M., Raimond, J. M. & Haroche, S. Rabi Oscillations Revival Induced by Time Reversal: A Test of Mesoscopic Quantum Coherence. *Phys. Rev. Lett.* **94**, 010401 (2005).

[9] Stupakov, G. V. Echo Effect In Hadron Colliders. *Ssc Report* SSCL- **579** (1992).

[10] Spentzouris, L. K., Ostiguy, J. -F. & Colestock, P. L. Direct Measurement of Diffusion Rates in High Energy Synchrotrons Using Longitudinal Beam Echoes. *Phys.Rev. Lett.* **76**, 620 (1996).

[11] Brown, R. W., Cheng, Y.-C. N., Haacke, E. M., Thompson, M. R. & Venkatesan, R. *Magnetic Resonance Imaging: Physical Principles and Sequence Design*, 2nd ed. (Wiley-Blackwell, 2014).

[12] Stupakov, G. Using the Beam-Echo Effect for Generation of Short-Wavelength Radiation. *Phys. Rev. Lett.* **102**, 074801 (2009).

[13] Xiang, D., Colby, E., Dunning, M., Gilevich, S., Hast, C., Jobe, K., McCormick, D., Nelson, J., Raubenheimer, T. O., Soong, K., Stupakov, G., Szalata, Z., Walz, D., Weathersby, S. & Woodley, M. Demonstration of the Echo-Enabled Harmonic Generation Technique for Short-Wavelength Seeded Free Electron Lasers. *Phys.Rev.Lett.* **105**, 114801 (2010).

[14] Zhao, Z. T., Wang, D., Chen, J. H., Chen, Z. H., Deng, H. X., Ding, J. G., Feng, C., Gu, Q., Huang, M. M., Lan, T. H., Leng, Y. B., Li, D. G., Lin, G. Q., Liu, B., Prat, E., Wang, X. T., Wang, Z. S., Ye, K. R., Yu, L. Y., Zhang, H. O., Zhang, J. Q., Zhang, Me., Zhang, Mi., Zhang, T., Zhong, S. P. & Zhou, Q. First lasing of an echo-enabled harmonic generation free-electron laser. *Nature Photonics* **6**, 360 (2012).

[15] Hemsing, E., Stupakov, G., Xiang, D. & Zholents, A. Beam by design: Laser manipulation of electrons in modern accelerators. *Rev. Mod. Phys.* **86**, 897 (2014).

[16] Parker J. & Stroud C. R. Jr. Coherence and decay of Rydberg wave Packets. *Phys. Rev. Lett.* **56**, 716 (1986).

[17] Averbukh, I. Sh. & Perelman, N. F. Fractional revivals: Universality in the long-term evolution of quantum wave packets beyond the correspondence principle dynamics. *Phys. Lett. A* **139**, 449-453 (1989).

[18] Robinett, R.W. Quantum wave packet revivals. *Physics Reports* **392**, pp.1-119 (2004).

[19] Karras, G., Hertz, E., Billard, F., Lavorel, B., Hartmann, J.-M., Faucher, O., Gershnabel, E., Prior, Y. & Averbukh, I. Sh. Orientation and Alignment Echoes. *Phys. Rev. Lett.* **114**, 15361 (2015).

[20] Karras, G., Hertz, E., Billard, F., Lavorel, B., Siour, G., Hartmann, J.-M., Faucher, O., Gershnabel, E., Prior, Y. & Averbukh, I. Sh. Fractional Echoes. arXiv:1603.01219 [physics.optics] (2016).

[21] Stapelfeldt, H. & Seideman, T. Aligning molecules with strong laser pulses. *Rev. Mod. Phys.* **75**, 543 (2003).

[22] Ohshima, Y. & Hasegawa, H. Coherent rotational excitation by intense nonresonant laser fields. *Int. Rev.Phys. Chem.* **29**, 619 (2010).

[23] Fleischer, S., Khodorkovsky, Y., Gershnabel, E., Prior, Y. & Averbukh, I. Sh. Molecular Alignment Induced by Ultrashort Laser Pulses and Its Impact on Molecular Motion. *Isr. J. Chem.* **52**, 414 (2012).

[24] Lemeshko, M., Krems, R. V., Doyle, J. M. & Kais, S. Manipulation of molecules with electromagnetic fields. *Mol. Phys.* **111**, 1648 (2013).

[25] Dörner, R., Mergel, V., Jagutzki, O., Spielberger, L., Ullrich, J., Moshammer, R. & Schmidt-Böcking, H. Cold Target Recoil Ion Momentum Spectroscopy: a "momentum microscope" to view atomic collision dynamics. *Physics Reports* **330**, 95-192 (2000).

[26] Lin, K., Song, Q. Y., Gong, X. C., Ji, Q. Y., Pan, H. F., Ding, J. X., Zeng, H. P. & Wu, J. Visualizing molecular unidirectional rotation. *Phys. Rev. A* **92**, 013410 (2015).

[27] Mizuse, K., Kitano, K., Hasegawa, H. & Ohshima, Y. Quantum unidirectional rotation directly imaged with molecules. *Sci. Adv.* **1**, e1400185 (2015).

[28] Boyd, R.W. *Nonlinear Optics* (Academic Press, Boston, 1992).

[29] Friedrich, B. & Herschbach, D. Alignment and Trapping of Molecules in Intense Laser Fields. *Phys. Rev. Lett.* **74**, 4623 (1995); Polarization of Molecules Induced by Intense Nonresonant Laser Fields. *J. Phys. Chem.* **99**, 15 686 (1995).

[30] Lichtenberg, A. J. & Lieberman, M. A. in *Regular and Chaotic Dynamics*, 2nd ed.; Marsden, J. E., Sirovich, L., Eds. *Applied Mathematical Sciences* Vol. 38. (Springer-Verlag: New York, 1992).

[31] Lynden-Bell, D. Statistical mechanics of violent relaxation in stellar systems. *Mon. Not. R. Astr. Soc.* **136**, 101 (1967).

[32] Lichtenberg, A. J. *Phase-space dynamics of particles* (Wiley New York, 1969).

[33] Dubetskii, B. Ya. & Chebotaev, V. P. Echoes in classical and quantum ensembles with determinate frequencies. *Pis'ma Zh.Eksp.Teor. Fiz.* **41**, No 6, 267-269 (1985).

[34] Dubetskii, B. Ya. & Chebotaev, V. P. Imaginary echo in a gas in a Doppler expanded transition. *Izvestiya Akademii Nauk SSSR, Seriya Fizicheskaya,* **50**, No 8, 1530-1536 (1986).

[35] Jiang, H. Y., Wu, C. Y., Zhang, H., Jiang, H. B., Yang, H. & Gong, Q. H. Alignment structures of rotational wavepacket created by two strong femtosecond laser pulses. *Opt. Express* **18**, 8990 (2010).

**Acknowledgements**

This work is supported by the National Natural Science Fund (11374103, 11434005 and 11322438), the Conseil Régional de Bourgogne (PARI program), the CNRS, the French National Research Agency (ANR) through the CoConicS program (contract ANR-13-BS08-0013), and the Labex ACTION program (contract ANR-11-LABX-0001-01). Support by the Israel Science Foundation (Grant No. 746/15), the ICORE program "Circle of Light" and the Minerva Foundation is highly appreciated. I.A. acknowledges support as the Patricia Elman Bildner Professorial Chair. This research was made possible in part by the historic generosity of the Harold Perlman Family.

**Author contributions**

Kang Lin and Peifen Lu performed the experiments; Kang Lin analyzed the data and wrote the manuscript; Junyang Ma, Xiaochun Gong, Qiying Song, Qinying Ji, Wenbin Zhang helped with the experiments and data analysis; Gabriel Karras, Guillaume Siour helped with the interpretation of the data; Jean-Michel Hartmann, Erez Gershnabel, Kang Lin and Ilya Averbukh performed the simulations; Ilya Averbukh, Yehiam Prior, Jian Wu, Heping Zeng and Olivier Faucher provided overall guidance to the project. All authors discussed the results and contributed to the manuscript.

**Competing financial interest**

The authors declare no competing financial interest.

# Supplementary Material for
# Molecular echoes in space and time

Kang Lin[1], Peifen Lu[1], Junyang Ma[1], Xiaochun Gong[1], Qiying
Song[1], Qinying Ji[1], Wenbin Zhang[1], Heping Zeng[1]*, and Jian Wu[1]†

[1]*State Key Laboratory of Precision Spectroscopy,
East China Normal University, Shanghai 200062, China*

Gabriel Karras[2], Guillaume Siour[3], Jean-Michel Hartmann[3], and Olivier Faucher[2]‡

[2]*Laboratoire Interdisciplinaire CARNOT de Bourgogne,
UMR 6303 CNRS-Université Bourgogne Franche-Comté, BP 47870, 21078 Dijon, France and*
[3]*Laboratoire Interuniversitaire des Systèmes Atmosphériques (LISA) CNRS (UMR 7583),
Université Paris Est Créteil, Université Paris Diderot, Institut Pierre-Simon Laplace,
Université Paris Est Créteil, 94010 Créteil Cedex, France*

Erez Gershnabel[4], Yehiam Prior[4]§, and Ilya Sh. Averbukh[4]¶

[4]*AMOS and Department of Chemical Physics,
Weizmann Institute of Science, Rehovot 76100, Israel*

(Dated: June 27, 2016)

**Quantum-mechanical simulation**

We model the alignment echo by numerically solving
the time-dependent Schroedinger equation $i\hbar\partial|\psi\rangle/\partial t = H_{\text{eff}}|\psi\rangle$ for the rotational state $|\psi\rangle = \sum_{\text{J,M}} C_{\text{JM}}|JM\rangle$,
where $H_{\text{eff}} = B_0 J(J+1) - 0.25\Delta\alpha\sin^2\theta(\varepsilon_z^2\cos^2\phi + \varepsilon_y^2\sin^2\phi + 2\varepsilon_z\varepsilon_y\cos\phi\sin\phi)$ is the effective Hamiltonian,
$B_0$ is the molecular rotational constant, $\Delta\alpha$ is the dif-
ference between the polarizability components parallel
and perpendicular to the molecular axis, $\theta$ and $\phi$ are
the polar and azimuth angles of the molecular axis with
respect to the X- and Z-axes, and $\varepsilon_y$ and $\varepsilon_z$ are the en-
velopes of the laser field vector along the Y- and Z-axes,
respectively. We first calculate the term $P_{J_0 M_0}(\theta,\phi,t) = |\sum_{\text{J,M}} C_{\text{JM}}(t)Y_{\text{JM}}(\theta,\phi)|^2$ for each initial molecular ro-
tational state $|\psi_{(t=0)}\rangle_{J_0 M_0} = |J_0, M_0\rangle$, where $Y_{\text{JM}}(\theta,\phi)$
is the spherical harmonic functions. We then assemble
them by considering the temperature-dependent Boltz-
mann distribution of the initial rotational states and the
proper spin statistics, and obtain the time-dependent
probability density distribution of the rotational wave
packet $P(\theta,\phi,t)$. For example, the molecular parameters
of $CO_2$ are: $B_0 = 0.39cm^{-1}$, $\Delta\alpha = 2.0\mathring{A}^3$. The rota-
tional temperature of the molecular beam is very close
to the translation temperature, which can be experimen-
tally estimated from $T_{\text{trans}} = \Delta p^2/[4ln(4)k_B m]$, where

$k_B$ is the Boltzmann constant, $\Delta p$ and $m$ are the full-
width at half-maximum (FWHM) of the momentum dis-
tribution (in the jet direction, i.e. y-axis) and mass of the
singly ionized $CO_2^+$, respectively. In our experiment we
measure a momentum width in the jet direction of $\Delta p \sim$
10.4 a.u. of $CO_2^+$ ions created by $P_1$ polarized along the
z-axis (orthogonally to the gas jet). The initial rotational
temperature of the $CO_2$ molecule is estimated to be 75K,
which shows good agreement with the experimental ob-
servations by matching the calculated arriving time of
the first alignment maximum as well as those of the frac-
tional revivals after the excitation of $P_1$. The intensities
of $P_1$ and $P_2$ are chosen to match the experiments well.

**Classical simulation**

We first initialize the rotational parameters of each
molecule $m$ of an ensemble of $N_m$ molecules, i.e. the
molecule axis orientation vector $\vec{u}_m(t=0)$ and angular
momentum $\vec{\omega}_m(t=0)$. This is done using Maxwell Boltz-
mann statistics at the chosen temperature for $||\vec{\omega}_m(t=0)||$, and random orientations for $\vec{u}_m(t=0)$ and $\vec{\omega}_m(t=0)$ [with $\vec{\omega}_m(t=0) \perp \vec{u}_m(t=0)$]. The molecules are then
free to rotate, a situation where $\vec{\omega}_m(t+dt) = \vec{\omega}_m(t)$ and
$\vec{u}_m(t+dt)$ is easily obtained from $\vec{u}_m(t)$ and $\vec{\omega}_m(t=0)$,
except during the laser pulses. In this latter case, we
first compute, for each molecule $m$, the polarizability ten-
sor $\overset{\leftrightarrow}{\alpha}_m^L(t)$ in the laboratory fixed frame from that $\overset{\leftrightarrow}{\alpha}_M$,
in the molecular frame: $\overset{\leftrightarrow}{\alpha}_m^L(t) = R_m(t) \cdot \overset{\leftrightarrow}{\alpha}_M \cdot R_m^{-1}(t)$,
where $R_m(t)$ is the rotation matrix connecting the two
frames [obtained from the orientation of the molecule
axis $\vec{u}_m(t)$]. The dipole $\vec{\mu}_m(t)$ induced in the molecule

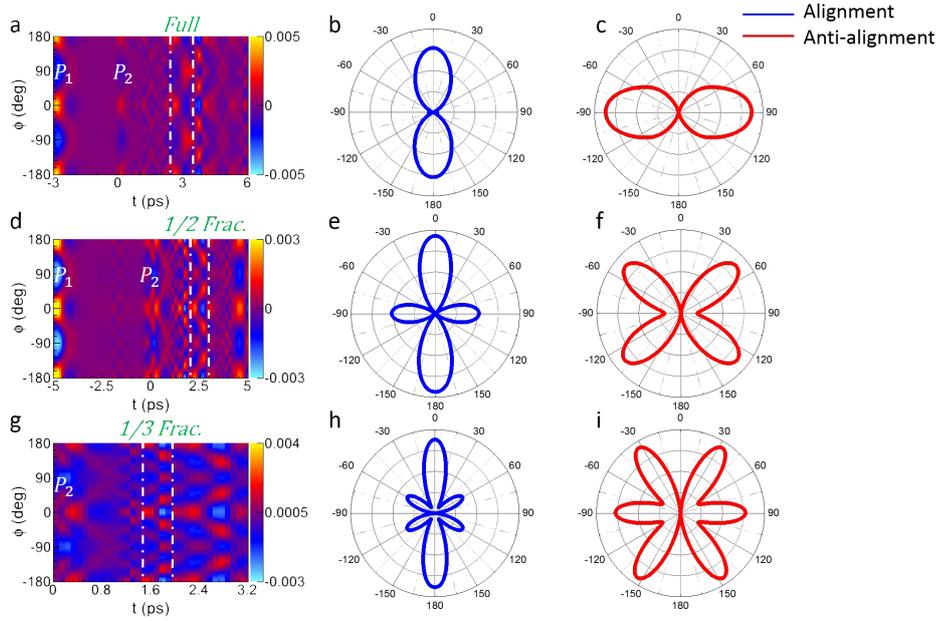

FIG. 1: **Quantum simulation of time-dependent angular distributions of full, 1/2 and 1/3 echoes.** (a) Angular distribution of full echo in case of parallel condition with a time delay of 3 ps between $P_1$ and $P_2$. (b) and (c) Alignment and anti-alignment of full echo in polar plots. (d)-(f) and (g)-(i) The same for 1/2 and 1/3 fractional echoes with a time delay of 5 ps between $P_1$ and $P_2$, respectively.

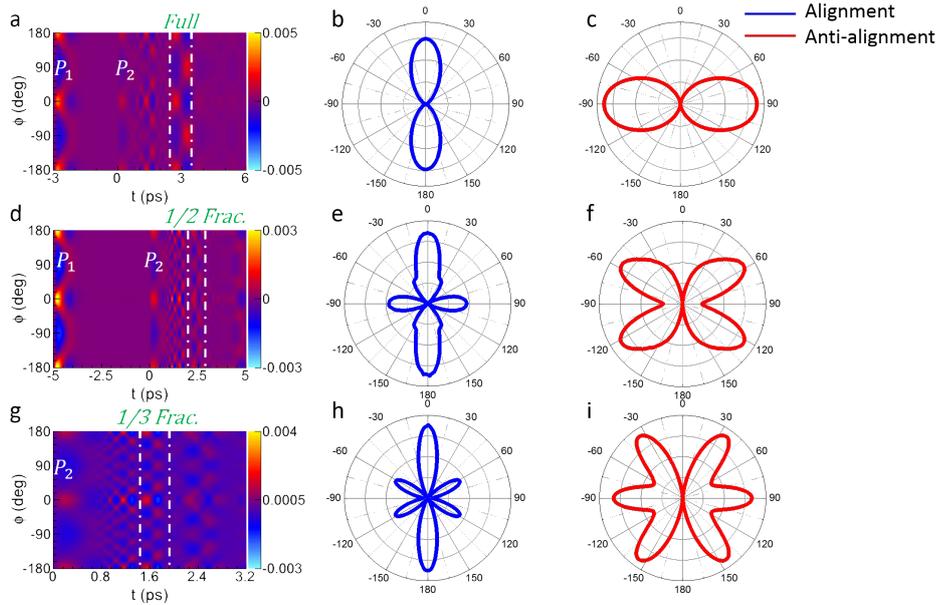

FIG. 2: **Classical simulation of time-dependent angular distributions of full, 1/2 and 1/3 echoes.** (a) Angular distribution of full echo in case of parallel condition with a time delay of 3 ps between $P_1$ and $P_2$. (b) and (c) Alignment and anti-alignment of full echo in polar plots. (d)-(f) and (g)-(i) The same for 1/2 and 1/3 fractional echoes with a time delay of 5 ps between $P_1$ and $P_2$, respectively.

by the laser field $\vec{E}(t)$ is then $\vec{\mu}_m(t) = \overleftrightarrow{\alpha}_m^L(t) \cdot \vec{E}(t)$, from which the torque $\vec{\tau}_m(t)$ subsequently applied to the molecule is computed from $\vec{\tau}_m(t) = \mu_m(t) \wedge \vec{E}(t)$. This enables the calculation of $d\vec{\omega}_m/dt(t) = \vec{\tau}_m(t)/I$, where $I$ is the molecule moment of inertia, from which

$\vec{\omega}_m(t+dt) = \vec{\omega}(t) + dt \times d\vec{\omega}_m/dt(t)$ is obtained enabling to deduce $\vec{u}_m(t+dt)$ from $\vec{u}_m(t)$, and so on... Note that, since the period of the laser field oscillation is much shorter than the laser pulse duration and the molecule rotational period, $|\vec{E}(t)|^2$ appearing in the expression of the torque can be replaced by $\varepsilon^2(t)/2$ where $\varepsilon(t)$ is the field envelop. Furthermore, in order to correctly sample the associated multidimensional phase space and obtain sufficiently converged results, several tens of millions of molecules must be used. A sufficiently small time step $dt$, with respect to the laser pulse(s) duration and molecular rotation period, but obviously be used.

**Supplementary Movie - rotated echo formation**

The movie depicts the time evolution of the molecular alignment angular distributions calculated using classical simulations for three different angles between the polarizations of two pump pulses.